# Revised research about chaotic dynamics in Manko *et al.* spacetime

Wen-biao Han[*]

*Shanghai Astronomical Observatory, Shanghai, China, 200030*

A recent work by Dubeibe *et al.* [Phys. Rev. D **75**, 023008 (2007)] stated that chaos phenomenon of test particles in gravitational field of rotating neutron stars which are described by Manko, Sanabria-Goméz, and Manko (Manko *et al.*) metric can only occur when the stars have oblate deformation. But the chaotic motions they found are limited in a very narrow zone which is very close to the center of the massive bodies. This paper argues that this is impossible because the region is actually inside of the stars, so the motions cannot exist at this place. In this paper, we scan all parameters space and find chaos and unstable fixed points outside of stars with big mass-quadrupole moments. The calculations show that chaos can only occur when the stars have prolate deformation. Because real deformation of stars should be oblate, all orbits of test particles around the rotating neutron stars described by Manko *et al.* solutions are regular. The case of nonzero dipolar magnetic moment has also been taken into account in this study.

## I. INTRODUCTION

Over the last decade or so, chaos in general relativity has been widely studied [1]. There are two main aspects in chaos dynamics in general relativity. One is the evolution of the gravitational field itself, such as the evolution dynamics of cosmology models, especially the mixmaster universe model [2–5]. The other is the dynamics of test particles in known spacetime. In the Newtonian frame, the history of study of chaos, complexity in astrophysics, and celestial mechanics, especially the *n*-body problem has spanned over 100 years from Poincaré. The role of chaos in stellar and galactic dynamics was also noticed [6]. Around the astrophysical objects with strong gravitational field, general relativity is a necessary tool to study motions of particles. Because general relativity is a strong nonlinear theory, chaos is more often visible. Especially, for some very interesting models, they are integrable in Newtonian theory but exhibit chaos in general relativity, such as the case of two fixed black holes [7] and the Schwarzschild black hole plus a dipolar shell [8,9].

In recent years, the spinning compact binaries dynamics has attracted great interest [10–19] because of the detecting project of gravitational-waves, such as LISA (Laser Interferometer Space Antenna), LIGO (Laser Interferometer Gravitational Wave Observatory), and ASTROD (Astrodynamical Space Test of Relativity using Optical Devices). The chaos phenomena of the binary systems has also sparked interest since the gravitational-waves detection cannot succeed when chaos is present.

Because it can be used to study dynamics evaluation of star clusters and galaxies, the study of orbital dynamics of test particles in known gravitational potential is significant.

In the astrophysical environment, many objects such as neutron stars or black holes are strong gravitational sources. The classical Newtonian theory is not satisfied any longer, so many people began to consider dynamics of test particles in known spacetime under the general relativity frame. As we know, the motions of nonspinning particles are integrable in Schwarzschild or Kerr spacetime. However, realistic astrophysical objects cannot be described very well by Schwarzschild or Kerr metric because of accretion discs or mass deformation existing. About the case of accretion discs, some authors have discussed it in many papers (Letelier *et al.* [20–23], Wu *et al.* [24], and references in them). In the present paper, we focus on the case of deformation and introduce it briefly as below.

Tomimatsu and Sato gave the most known Tomimatsu-Sato family (hereafter TS-family) solution [25,26], in which quadrupole deformation in the nonrotating limit is $\mathcal{Q} = -\frac{2}{\delta^3}m^3$, being $m = \frac{\delta M_0}{2}$ where $M_0$ is the mass monopole of the source and $\delta$ a dimensionless parameter, taking the $\delta = 1$ for the Kerr solution and $\delta = 2$ for the Tomimatsu-Sato $\delta = 2$ solution (henceforth TS2 solution). They [26] and then Tanabe [27] pointed out that TS-family represents the gravitational field of rotating masses with angular momentum about the z-axis. Guéron and Letelier [28] analyzed the influence of the introduction of multipolar terms corresponding to the quadrupole deformation in the Schwarzschild and Kerr solutions finding chaotic behavior for some values of the deformation parameter. In 2000, Manko, Mielke, and Sanabria-Goméz [29] and Manko, Sanabria-Goméz, and Manko [30] generalized a new memer of the vacuum TS-family, an asymptotically-flat solution to the Ernst equations for the Einstein-Maxwell system (hereafter Manko *et al.* solution). The Manko *et al.* spacetime depends on five parameters: electric charge $Q$, magnetic dipole $\mu$, the gravitational mass $m$, the angular momentum $a$, and an arbitrary quadrupolar deformation $b$. Besides, Berti and Stergioulas [31], Berti, White, Maniopoulou, and Bruni [32] studied this solution as a possible model to describe the gravitational field of a rapidly rotating neutron stars (hereafter RRNS). They matched the Manko *et al.* solutions with numerical relativity results, and pointed out that the quadrupole moments of them are the same. But, only one of the two branches of analytic solutions ($b$-branch) matches the numerical solutions for rapidly rotating neutron stars. Then, Dubeibe and Pachón studied the dynamics of test particles in the T-S and Manko *et al.* solutions, and they found that only the last one can exhibit chaos in some very narrow numerical instable domain when the RRNS have oblate deformation ($\mathcal{Q} < 0$) [33].

---

[*]wbhan@shao.ac.cn
Also at Graduate School of the Chinese Academy of Sciences, Beijing, China.

In this paper, we follow Dubeibe and Pachón [33] to scan all parameters to study the dynamical model of the above mentioned. We argue that the chaotic solutions found by Dubeibe and Pachón of the oblate deformation neutron stars are unpractical in fact. It is because these zones are inside of the central objects. So the chaos cases found by Dubeibe *et al.* cannot exist in realistic astrophysical conditions. And also, for oblate deformation, we scan all physical parameters space and find that orbits of test particles are nonchaotic except the cases of Dubeibe *et al.* But for larger quadrupole deformation related parameter $b$, we find new chaotic regions and unstable fixed points outside of the massive central bodies which have prolate deformation and physical rotating parameters $a$. This opinion was not mentioned by Dubeibe and Pachón in their paper [33]. Considering astrophysical neutron stars and black holes are neutral and strongly magnetic, we also choose $Q = 0$ but $\mu \neq 0$ in some cases and simply study the influence of dipolar magnetic moment to the dynamics of our systems.

The paper is organized as follows. In Sec. II we introduce the Manko *et al.* solution and discuss it simply. Then, in Sec. III we give the geodesic dynamics equations for test particles in Manko *et al.* spacetime. In Sec. IV we investigate the chaos dynamics of the above systems. Two cases are considered, one is $\mu = 0$, the other nonzero. At last, in Sec. V we make some brief conclusions and discussions.

Throughout the work we use geometric units $c = G = 1$, and take the signature of a metric as $(-, +, +, +)$. Greek subscripts run from 0 to 3, and Latin indexes run from 1 to 3.

## II. GRAVITATIONAL FIELD OF A RAPIDLY ROTATING STAR

In the vacuum region, the most simple form of the spacetime around a stationary and axisymmetric body was given by Papapetrou [34]. It can be written as

$$ds^2 = -f(dt - \omega d\phi)^2 + f^{-1}[e^{2\gamma}(d\rho^2 + dz^2) + \rho^2 d\phi^2], \quad (1)$$

where $f$, $\omega$, and $\gamma$ are only functions of the quasicylindrical Weyl coordinates $(\rho, z)$. The Wyle coordinates are related with the prolate spheroidal coordinates $(x, y)$ by means of the transformation

$$\rho^2 = \kappa^2(x^2 - 1)(1 - y^2), \qquad z = \kappa xy \quad (2)$$

where $x \geq 1$, $-1 \leq y \leq 1$ and $\kappa$ is a parameter which will be explained later.

After more than ten years of work in the theory of fields, Manko *et al.* found a vacuum solution involving five parameters: mass, angular momentum, charge, magnetic dipole moment, and quadrupole moment [30]. Considering all realistic bodies are neutral, so, in our cases, the charge $Q \equiv 0$. Thus, the quadrupolar deformation of Manko *et al.* solution is written as

$$\mathcal{Q} = M[\delta - d - a(a - b)]. \quad (3)$$

with

$$\delta = \frac{\mu^2 - M^2 b^2}{M^2 - (a - b)^2}, \qquad d = \frac{1}{4}[M^2 - (a - b)^2], \quad (4)$$

where $M$ denotes gravitational mass of the central body, $a(= J/M)$ the specific angular momentum, $b$ a parameter related with the mass-quadrupole moment, and $\mu$ a parameter that can be related to the dipolar magnetic moment. For $Q = 0$, the metric functions take the form,

$$f = \frac{E}{D}, \qquad \omega = \frac{-(1 - y^2)F}{E} \qquad e^{2\gamma} = \frac{E}{16\kappa^8(x^2 - y^2)^4} \quad (5)$$

with

$$D = \{4(\kappa^2 x^2 - \delta y^2)^2 + 2\kappa Mx[2\kappa^2(x^2 - 1) + (2\delta + ab - b^2(1 - y^2)] \\
+ (a - b)[(a - b)(d - \delta) - M^2 b](y^4 - 1) - 4d^2\}^2 + 4y^2\{2\kappa^2(x^2 - 1)[\kappa x(a - b) - Mb] - 2Mb\delta(1 - y^2) \\
+ [(a - b)(d - \delta) - M^2 b](2\kappa x + M)(1 - y^2)\}^2 \quad (6)$$

$$E = \{4[\kappa^2(x^2 - 1) + \delta(1 - y^2)]^2 + (a - b)[(a - b)(d - \delta) - M^2 b](1 - y^2)^2\}^2 \\
- 16\kappa^2(x^2 - 1)(1 - y^2)\{(a - b)[\kappa^2(x^2 - y^2) + 2\delta y^2] + M^2 b y^2\}^2 \quad (7)$$

$$F = 8\kappa^2(x^2-1)\{(a-b)[\kappa^2(x^2-y^2)+2\delta y^2]+y^2M^2b\}\{\kappa Mx[(2\kappa x+M)^2-a^2+b^2-2y^2(2\delta+ab-b^2)]$$
$$-2y^2(4\delta d-M^2b^2)\}+\{4[\kappa^2(x^2-1)+\delta(1-y^2)]^2+(a-b)[(a-b)(d-\delta)-M^2b](1-y^2)^2\}$$
$$\times(4(2\kappa Mbx+2M^2b)[\kappa^2(x^2-1)+\delta(1-y^2)]+(1-y^2)\{(a-b)(M^2b^2-4\delta d)$$
$$-(4\kappa Mx+2M^2)[(a-b)(d-\delta)-M^2b]\}), \tag{8}$$

where $x = \frac{1}{2\kappa}(r_+ + r_-)$ and $y = \frac{1}{2\kappa}(r_+ - r_-)$ with $\kappa = \sqrt{d+\delta}$ and $r_\pm = \sqrt{\rho^2 + (z\pm\kappa)^2}$.

Now, we consider the quadrupolar moment $Q$. From Eq. (3), $Q$ depends on $M$, $a$, $b$, and $\mu$. For simplicity, we set $\mu = 0$ at first. If choosing a group of typical data given by Berti and Stergioulas [31]: $M = 2.904$, $a = 1.549$, we can give a relation of $Q$ about $b$. This is shown in Fig. 1. For the oblate solutions ($Q < 0$), there is a minimum quadrupolar deformation $|Q| = 10.875$ while $b = 0.466$. And for prolate deformation, there are two minimum points $(-2.851, 6.358)$ and $(7.032, 96.355)$.

## III. DYNAMICS EQUATIONS FOR TEST PARTICLES

For nonspining free test particles, the geodesic equation is

$$\frac{d^2x^\lambda}{d\tau^2} = -\Gamma^\lambda_{\mu\nu}\frac{dx^\mu}{d\tau}\frac{dx^\nu}{d\tau}, \tag{9}$$

where, $\tau$ is defined as proper time and $\Gamma^\lambda_{\mu\nu}$ is the Christoffel symbol. And it should be noted that the $x$ here represents coordinate variables $(t, \rho, z, \phi)$. The geodesic motion has an important character, namely, for each symmetry of the background spacetime (typically represented by a Killing vector) there is a corresponding a constant of motion. Obviously there are two Killing vectors $\vec{\xi}^t = \partial/\partial t$ and $\vec{\xi}^\phi = \partial/\partial\phi$ for Manko et al. solution. Then they give energy and $z$ angular momentum conservation:

$$E = -p_t = f(\dot{t} - \omega\dot{\phi}) \tag{10}$$

and

$$L = p_\phi = \omega f(\dot{t} - \omega\dot{\phi}) - \frac{\rho^2}{f}\dot{\phi}, \tag{11}$$

where the constants of integration $E$ and $L$ denote the energy and the angular momentum of the test particles, respectively. So, for the Manko et al. spacetime, Eq. (9) can be written as a four-dimensional dynamics system in the variables $(\rho, z, \dot{\rho}, \dot{z})$. And the dynamics equations can be cast as

$$\ddot{\rho} = \frac{g_{ab,\rho}\dot{x}^a\dot{x}^b + g_{zz,\rho}\dot{z}^2 - g_{\rho\rho,\rho}\dot{\rho}^2 - 2g_{\rho\rho,z}\dot{\rho}\dot{z}}{g_{\rho\rho}} \tag{12}$$

$$\ddot{z} = \frac{g_{ab,z}\dot{x}^a\dot{x}^b + g_{\rho\rho,z}\dot{\rho}^2 - g_{zz,z}\dot{z}^2 - 2g_{zz,\rho}\dot{\rho}\dot{z}}{g_{zz}}, \tag{13}$$

where the overdots denote derivation with respect to the proper time $\tau$ and the indices $a$ and $b$ take the values $(t, \phi)$, the metric function $g_{\mu\nu}$ is given by Eqs. (1) and (4)–(8).

The system admits a third integration constant, named four-velocity conservation

$$g_{\mu\nu}\dot{x}^\mu\dot{x}^\nu = -1. \tag{14}$$

Now, we introduce the process of our numerical integration method. First, we give energy $E$ and angular momentum $L$, then set initial conditions: $t_0 = 0$, $\rho_0$, $z_0$, and $\phi_0$ So that the $\dot{t}$ and $\dot{\phi}$ can be calculated by Eqs. (10) and (11). If one of the $\dot{\rho}_0$ or $\dot{z}_0$ is given out, according to Eq. (14), the other one can be calculated out immediately. At last, we numerically integrate Eqs. (12) and (13), and check numerical precision by the four-velocity conservation (14). The adopted integrator is RKF7(8), and errors of Eq. (14) are about $10^{-13}$–$10^{-14}$ usually.

## IV. CHAOS DYNAMICS OF MANKO'S SOLUTION

The qualitative aspects of the system (12) and (13) can be studied conveniently to compute the Poincaré sections through the plane $z = 0$.

First, we discuss the chaos solutions found by Dubeibe et al. [33]. They modified sightly the values of the parameter $b$ given by [31] to change the quadrupole deformation and found chaotic behavior of the system. They gave two groups parameters in their paper: $M = 2.904$, $a = 1.549$, $b = 0.8$, $E = 0.96$, $L = 9.9$, and $E = 0.971$, $L = 9.3$. Both of them are oblate deformation. We give the Poincaré sections of the first group parameters in Fig. 2, and can find chaotic motion on the left-hand side. It is very clear that the distance of chaotic trajectories on the left-hand side in Fig. 2 is less than the Schwarzschild radii $2M(5.808)$ of the star, so these orbits are actually inside of the central body. Results of the other chaotic solutions they found are the same as in Fig. 2. So, we think the chaotic solutions found by Dubeibe et al. could not exist in fact. Furthermore, for oblate deformation cases, we scan all parameter space and find no other chaotic solutions, except the kind of unphysical chaotic solutions mentioned above. Accordingly, we think that chaotic motions of test particles around the oblate neutron stars described by Manko et al. spacetime cannot occur in fact.

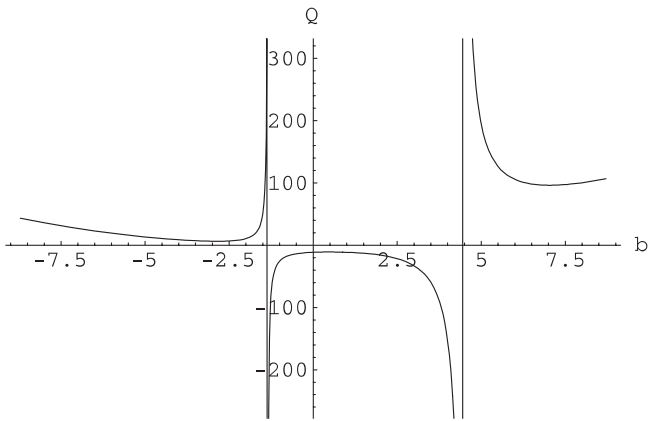

FIG. 1. $Q$ as a function of $b$ while $M = 2.904$, $a = 1.549$. The mass deformation is oblate when $-1.355 < b < 4.453$, and for other $b$, is prolate. There are three minimum quadrupolar deformation values while $b = 0.466$, $-2.85125$, and $7.032$.

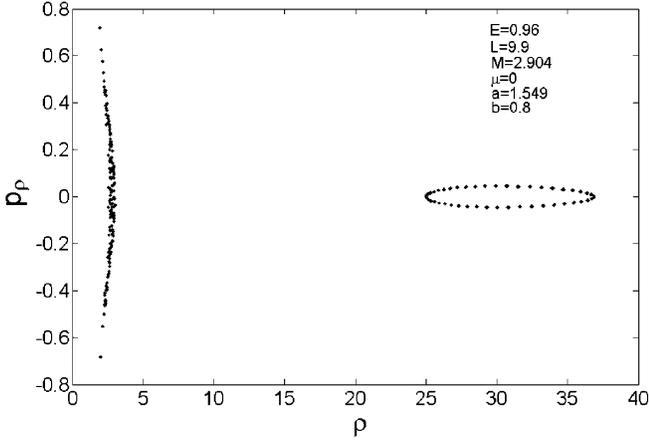

FIG. 2. The Poincaré sections of a set of parameters with oblate deformation. It has two apart regions, one is the narrow chaotic zone on the left-hand side, the other is the wide regular motion zone. But the chaotic orbits in fact are inside of the central body.

Now, we consider the prolate deformation solutions. In the Ref. [33], the authors did not discuss the cases of prolate deformation, but they thought that chaos can only occur in the oblate case. Naturally, we doubt this point. From Fig. 1, when choosing $b > 4.453$, then the $Q > 0$, and the solutions are prolate. We choose $b = 6.0$ and angular momentum $L = 2.75M$, and $E$, $a$, $\mu$ are the same with Fig. 2. We find strong chaotic motions at this set of parameters, and show the Pioncaré sections in Fig. 3. Obviously, the chaotic region in Fig. 3 is far away from the Schwarzschild radii of central star. Furthermore, according to data given by Berti [31], the equatorial circumferential radius of the neutron star with this mass is 11.62 km, and the minimum distance of the chaos orbit from the center of body is 11.712 km in Fig. 3. Considering the prolate case, the equatorial circumferential radius should be smaller. So, for prolate neutron stars without magnetic field, our calculations indicate that the chaotic geodesic motions around it can exist.

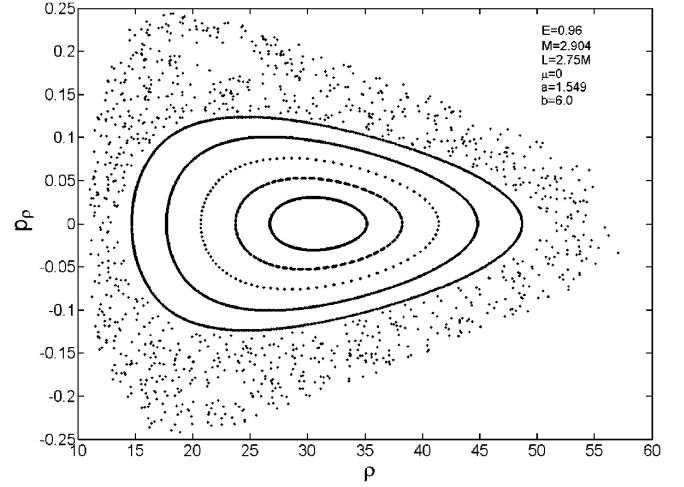

FIG. 3. The Poincaré sections of a set of parameters with prolate deformation. It clearly shows chaos at the exterior of the central star (black hole or neutron star).

More calculations show that when $b$ values ($b > 0$) around the neighborhood of the minimum prolate deformation, chaotic motions can also be found. But for the other branch of prolate deformation ($b < 0$), we still cannot find chaos solutions.

Additionally, the dynamical systems present hyperbolic equilibrium points for some special $b$ values before global chaos emerges. For example, the Fig. 4 gives the Poincaré sections of a set of parameters: $E = 0.95$, $L = 3.0$, $a = 0.6$, and $b = 3.0$ with unit mass. It shows that the system presents a hyperbolic equilibrium point at (8.65, 0) and three little loops belonging to the same trajectories called a chain of islands [35]. In Fig. 5, using the invariant fast Lyapulov indicator(FLI) defined by Wu et al. [36], we find the hyperbolic equilibrium point is chaotic too.

So, based on the above discussion, for Manko et al. spacetime without dipolar magnetic moment, chaotic motions of the oblate deformation cases cannot exist in fact,

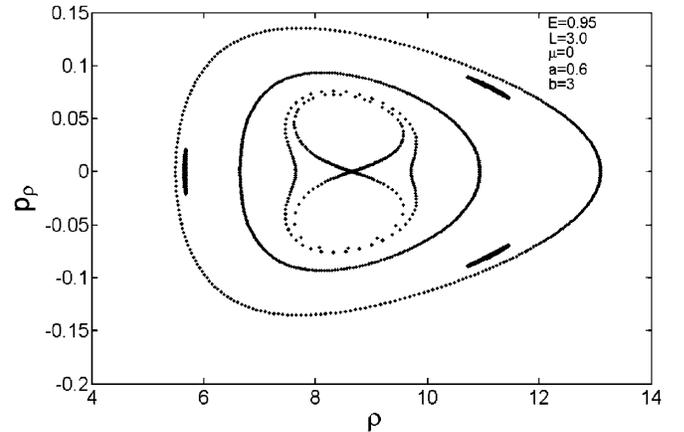

FIG. 4. An unstable fix point and a chain of islands at parameters: $E = 0.95$, $L = 3.0$, $a = 0.6$, and $b = 3.0$ with unit mass.

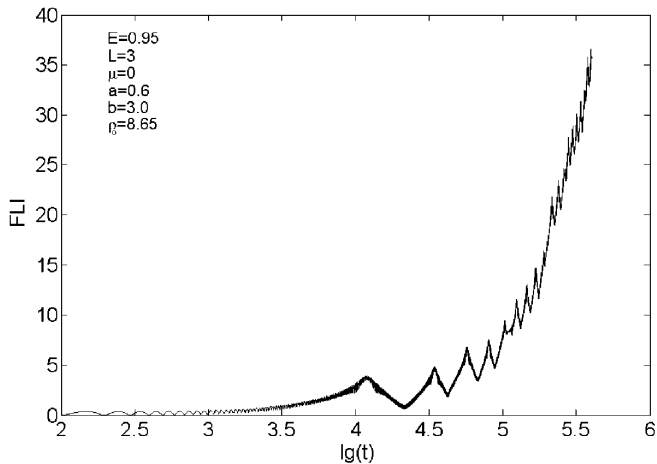

FIG. 5. The FLI of the hyperbolic equilibrium point in Fig. 4. An exponential raise of FLI presents it as a chaotic motion.

but the prolate cases can permit chaotic orbits or complicated geodesic motion.

Additionally, we study the cases of nonzero dipolar magnetic moment $\mu$. A nonzero $\mu$ cannot change the global dynamics character of the system (10)–(13), because the metric functions (5)–(8) do not include $\mu$ explicitly. In general, the astrophysical objects are oblate, so we suspect the oblate bodies with magnetic moment can produce chaotic solutions. But our calculations indicate that there is not chaos around the oblate stars with dipolar magnetic moment. The gravitational fields of the model are described by Manko et al. solutions.

## V. CONCLUSIONS

In this paper, we revisit the chaos dynamics of test particles moving in Manko et al. spacetime, and find some different and interesting phenomena. We think it is impossible to present chaos regions for the oblate deformation massive bodies, because those regions found by Dubeibe et al. are actually inside neutron stars. However, for the prolate one we find chaotic solutions for some values of $b$ parameter. Furthermore, we do not think that the nonzero dipolar magnetic moment can change the dynamical structure of the systems. For realistic astrophysical objects that are oblate, the present paper indicates that there are not chaotic solutions of test particles around single oblate deformation neutron stars described by Manko et al. solutions.

The phenomenon of chaos should be a physical existence, whether in Newtonian theory or general relativity [37]. But in general relativity, sometimes the classical Lyapunov exponents (LE hereafter) are noninvariant, because of the choice of spacetime coordinates. So, if using classical LE to detect chaos, it looks like the chaos depends on the spacetime parametrizations. In this paper we mainly use the Poincaré sections to detect chaos; this is independent of spacetime coordinates. Additionally, the relativistic invariant fast Lyapunov indicator (FLI) defined by Wu et al. [36] is used in Fig. 5. So our conclusions do not depend on spacetime parametrizations.

In Manko et al. solution, the magnetic field does not change the spacetime essentially, and it is unlikely in astrophysics. Very strong magnetic fields are universal around black holes or neutron stars, and should have important influence on the gravitational field. So we need to study the dynamics of particles around more realistic astrophysical models. Perhaps this should be considered in our future work.

In the first section, we mentioned the problem of chaos in spinning compact binary systems, for which the existence of chaos remains controversial (see Ref. [12] and subsequent discussion after it by Schnittman, Rasio, Cornish, and Levin et al.). The models of test particles in known spacetime are the most ideal simplification of binaries. The models are one-body problems, not two-body problems. To date, because of complication of the two-body problem, the research tool of chaos dynamics is still the post-Newtonian approximation, not general relativity. However, if we find chaos in the ideal one-body models [38], in more complex binary systems chaotic motions should be unavoidable.

On the other hand, what roles can chaos dynamics play in astronomy? Many people think it plays an important role in astrophysics and dynamical astronomy, such as the evolution of tidal-capture binaries, stellar or galactic dynamics, the evolution of $n$-body systems and so on [6,39,40]. Perhaps, the research on how chaos plays a role in astrophysics is quite significant.

## ACKNOWLEDGMENTS

The author thanks for the help of Professor Xinhao Liao and Xin Wu. Miss Wei Ma and Professor Liguang Li help me to improve the presentation.